\documentclass[conference]{IEEEtran}
\pdfoutput=1
\usepackage[hyphens]{url}
\usepackage{hyperref}
\hypersetup{breaklinks=true, colorlinks,allcolors=blue}
\usepackage[backend=biber, style=ieee]{biblatex}

\usepackage{amsmath,amssymb,amsfonts}
\usepackage{algorithmic}
\usepackage{graphicx}
\usepackage{textcomp}
\usepackage{xcolor}
\usepackage{orcidlink}  

\def\BibTeX{{\rm B\kern-.05em{\sc i\kern-.025em b}\kern-.08em
    T\kern-.1667em\lower.7ex\hbox{E}\kern-.125emX}}
\begin{document}

\title{The Psychological Impacts of Algorithmic and AI-Driven Social Media on Teenagers:\\ A Call to Action
}

\author{\IEEEauthorblockN{Sunil Arora\,\orcidlink{0009-0007-3066-3461}}
\IEEEauthorblockA{\textit{Dakota State University}, USA \\
sunil.arora@trojans.dsu.edu}
\and
\IEEEauthorblockN{Sahil Arora\,\orcidlink{0009-0004-6138-6135}}
\IEEEauthorblockA{
sahil.arora3117@gmail.com}
\and
\IEEEauthorblockN{John D. Hastings\,\orcidlink{0000-0003-0871-3622}}
\IEEEauthorblockA{\textit{Dakota State University}, USA \\
john.hastings@dsu.edu}
}

\maketitle

\begin{abstract}
This study investigates the meta-issues surrounding social media, which, while theoretically designed to enhance social interactions and improve our social lives by facilitating the sharing of personal experiences and life events, often results in adverse psychological impacts. Our investigation reveals a paradoxical outcome: rather than fostering closer relationships and improving social lives, the algorithms and structures that underlie  social media platforms inadvertently contribute to a profound psychological impact on individuals, influencing them in unforeseen ways. This phenomenon is particularly pronounced among teenagers, who are disproportionately affected by curated online personas, peer pressure to present a perfect digital image, and the constant bombardment of notifications and updates that characterize their social media experience. As such, we issue a call to action for policymakers, platform developers, and educators to prioritize the well-being of teenagers in the digital age and work towards creating secure and safe social media platforms that protect the young from harm, online harassment, and exploitation.
\end{abstract}


\section{Introduction}

Social media is a persuasive technology, a concept introduced by B.J. Fogg ~\cite{fogg1998} in the late 1990s. Persuasive technologies are designed to interactively 
influence people's attitudes or behaviors. Social media platforms use specialized algorithms, AI models, and user interfaces designed to capture and retain user attention, using techniques such as personalized recommendation content feeds, notifications, and interactive designs. Such methods aim to maximize user engagement, often leading to prolonged and repetitive use. 

On January 31, 2024, the United States Senate Judiciary Committee summoned social media companies' Chief Executive Officers (CEOs), including Meta, TikTok, X, Discord, and Snapchat, for failure to protect children from harm and child sexual abuse on social media platforms \cite{senatejudiciary}. Several parents of children, including those whose children died due to online extortion, sexual abuse, and exposure to harmful content on social media platforms, attended the hearing \cite{wsj}.

The current social media landscape prioritizes sensationalism and engagement over factual accuracy. It creates a culture where individuals feel compelled to curate a polished, relatable, and entertaining online persona while navigating the treacherous waters of public judgment. The result is a culture of superficial connections and shallow thinking, where the art of deep conversation and critical thinking gradually erodes.

The pervasive use of social media by children raises significant concerns about its impact on their mental and physical health. Research indicates that young users are particularly susceptible to the persuasive elements of social media \cite{pew, pulsesurvey1}. Continuous and endless scrolling, facilitated by infinite content feeds, can lead to excessive screen time, which has been associated with a range of adverse psychological outcomes \cite{OASH}.  These risks highlight the importance of continuous research, discussions, and efforts by the research community, technology industry, and governments to protect children from social media's psychological impacts, harm, and abuse.

This paper explores the challenges associated with social media use by children, along with potential solutions, guided by the following research questions:
\begin{itemize}
    \item What factors contribute to the psychological impacts of algorithmic and AI-driven social media on teenagers?
    \item How can legislators, social media providers, and educators address these challenges to minimize the impact of social media?
    \item What are the potential benefits and limitations of alternative social media models in addressing these issues?
\end{itemize}

The remainder of the paper is organized as follows. Section \ref{Psych} discusses social media's psychological impact. Section \ref{DoomScroll} presents the current meta-issues with social media and content consumption. Section \ref{WhatCanBeDone} details the necessary actions to be taken by governments, educational institutions, and the social media industry, followed by the positive potential of Mastodon in Section \ref{Mastodon}.  Section \ref{CallToAction} outlines actionable recommendations and Section \ref{Conclusion} concludes the paper.

\section{Psychological Impacts of Social Media on Teens}\label{Psych}

The widespread use of social media by children and its effect on their mental health has garnered significant attention from the media, researchers, and governments recently. Several studies emphasize the connection between social media usage and psychological problems among children and teenagers \cite{mougharbel2023heavy, keles}. 

U.S. Surgeon General Dr. Vivek Murthy issued an advisory in May 2023 highlighting the risk social media poses to children's mental health and well-being \cite{OASH}. According to a 2023 survey, nearly half of the 1453 surveyed children between the ages of 13 and 17 in the United States used social media apps almost constantly, which doubled compared to 2014-2015~\cite{pew}. The other half reported using social media several times a day. Another survey of 1480 children between 13 and 17, conducted by Boston Children Digital Wellness Lab in 2022, reported that children spent an average of 8.2 hours daily on social media, and 57\% felt they use it too much~\cite{pulsesurvey1}. In another survey by Boston Children Digital Wellness Lab in 2023 \cite{pulsesurvey2}, 38\% of the children reported a negative (e.g., uncomfortable,  unsafe) experience with social media platforms. 

A systematic review conducted by \citeauthor{keles}~\cite{keles} indicated a noteworthy association between extensive social media usage and its impact, including psychological distress, anxiety, and depressive symptoms among teenagers. The review highlighted factors such as the constant need for social validation and exposure to cyberbullying as contributing factors \cite{keles}. The harmful effects of social media can extend to severe outcomes such as suicidal thoughts and behavior. Multiple studies have documented cases where excessive use of social media, screen time, and online victimization have led to tragic outcomes among adolescents \cite{CHEN, kim, senkal, peprah}.

These findings highlight the immediate need for action and effective strategies to mitigate social media's negative psychological impacts on children.

\section{The Rise of Doom Scrolling: A Social Media Phenomenon}\label{DoomScroll}
Doom scrolling refers to the act of mindlessly scrolling through social media, often out of boredom, habit, or anxiety. For many individuals, their smartphones have become a trusted companion, offering a welcome respite from the demands of daily life. This reliance on mobile devices was particularly pronounced during the COVID-19 pandemic, as people sought solace in their phones during extended periods of isolation and remote work \cite{Grosser2022wd, Sharma2022}.

According to Pew Research Center findings~\cite{Vogels_2023}, YouTube is the most frequently used social media platform among teenagers, with an overwhelming majority (77\%) using it every day. While TikTok usage is not as widespread, still a significant proportion (58\%) of teens engage with this app daily. Instagram and Snapchat are also popular choices for daily use, with roughly half (50\% and 51\%, respectively) of teens reporting they use these platforms at least once a day. Interestingly, only about one in five teens (19\%) say they use Facebook daily. 

Short-form video content has become increasingly popular, with TikTok, Snapchat, and YouTube emerging as the leading platforms for sharing bite-sized videos. Research suggests that prolonged consumption of short-form videos can lead to difficulties in concentration, information retention, and a preference for instant gratification over longer content, ultimately affecting attention span and academic focus \cite{Asif_Saniya_Kazi_2024}. As we mindlessly scroll through social media feeds, the constant influx of short-form videos can create a sense of FOMO (fear of missing out) that drives us to keep watching. But this addiction to instant entertainment comes at a cost, as our brains become conditioned to crave quick hits of dopamine rather than engage in more meaningful activities. 

Humans have a natural limitation when it comes to focused attention. Our brains can only maintain concentration for so long before we begin to lose momentum. This capacity for sustained focus is not set in stone; various factors, such as the complexity of the task, personal interest, motivation levels, and individual experiences, can influence it \cite{Shanmugasundaram_2023}.

\section{What Can Be Done?}\label{WhatCanBeDone}
The impact of social media on teenagers' mental health is a pressing concern that requires immediate attention. A comprehensive approach must be considered to address this issue, involving government regulations, industry-led initiatives, and educational programs.

\subsection{Government Regulations}
Governments and policymakers have obligations to regulate emerging technologies, but they face scrutiny and pushback from the technology industry on their decisions~\cite{johnson2023doctrine}. At the same time, technology companies are criticized for their technology designs and their social and political impacts. Technology providers are often called to proactively consider the ramifications of their decisions early in the design process. However, they may not understand or be unwilling to respond to  social issues, thus people demand their political leaders take necessary actions \cite{wolff}. 

In the last few years, governments worldwide have actively introduced regulations to protect their citizens, especially children, from social media harm. Regulations and legislation can be critical in protecting children from social media harm. Legislation serves as the foundation for establishing a legal framework and allocating resources to address instances of misconduct. Well-crafted legislation enables governments to influence how social media platforms operate, particularly for the younger demographic \cite{Bacon2023}. 

In the United States, federal and state governments have introduced and enacted various legislation to protect children from social media harm. Table \ref{tab:tab1} outlines the recent legislation enacted or introduced at the state and federal levels in addition to other existing legislation, such as the Children's Online Privacy Protection Act (COPPA)~\cite{coppa1998}. Section 230~\cite{section230}, an amendment to the Communications Act of 1934 \cite{commact1934} and enacted as part of the Communications Decency Act of 1996, is a significant topic in any discussion about U.S. legislation related to social media. Section 230 grants immunity to social media providers and users, and consists of two key parts. Section 230(c)(1) outlines that service providers or users cannot be considered the publisher of any information provided by other users, and social media service providers or users cannot be held liable for acting in good faith to limit access to certain types of objectionable material. \cite{jacobs2024, zhang2024, crs}.

On April 30, 2024, Senator Brian Schatz introduced new federal legislation, the `Kids Off Social Media Act'~\cite{UScongress}. This bill prohibits the use of social media by children under 13 and personalized recommendations for those between the ages of 12 and 17. In addition, it requires social media platform companies to terminate social media accounts of ages under 13 and delete their personal data \cite{UScongress, senatec}. The new legislation also includes the Eyes on the Board Act of 2023~\cite{US2023}, which instructs schools to restrict and monitor the use of social media for kids on school devices and networks and implement a screen time policy. The Earn IT Act of 2023 \cite{earnitact} establishes a National Commission on Online Child Sexual Exploitation Prevention. This bill limits the liability protections of social media providers concerning claims related to child sexual exploitation. In addition, the Earn IT Act of 2023 changes the reporting requirements for social media providers for child sexual exploitation reports to the National Center for Missing and Exploited Children, including technical facts reports and preserving the content for more than one year.  The Strengthening Transparency and Obligations to Protect Children Suffering from Abuse and Mistreatment (STOP CSAM) Act of 2023 \cite{stopcasm} mandates child abuse reporting, expands protections for child victims, empowers victims to request removal of child sexual abuse material from tech platforms, and holds tech companies accountable and class action for promoting or facilitating online child sexual exploitation. Additionally, it strengthens CyberTipline reporting requirements and mandates social media companies to submit annual reports on promoting a culture of safety for children on their platforms \cite{stopcasm}.

In 2024, the Utah government passed two pieces of legislation, the Utah Minor Protection in Social Media Act (S.B. 194)~\cite{utah194} and the Utah Social Media Amendments (H.B. 464)~\cite{utah464}, which will be enacted in October 2024 to protect from social media harms and hold social media companies accountable for any mental health issues due to algorithmically curated social media service.

Similarly, New York state passed legislation, Senate Bill S7694~\cite{nysenate}, called the Stop Addictive Feeds Exploitation (SAFE) Act, to protect children from addictive feeds, provide a mechanism for parents to control their usage, and require parental consent for notifications between 12 AM and 6 AM. It highlights the importance of protecting our young generation from social media harms. These efforts  by the U.S. government also acknowledge the harms and effects of social media on children. 
\begin{table}[h!]
    \centering
\caption{Recently Introduced or Passed  U.S. Social Media Legislation}
\label{tab:tab1}
\begin{tabular}{|p{2cm}|p{4cm}|p{1.5cm}|}
\hline 
         \textbf{U.S. Legislation}&   \textbf{Purpose}&\textbf{Status}\\ \hline 
STOP CSAM Act of 2023\cite{stopcasm}&   Protects children from sexual exploitation and promotes accountability and transparency in the technology industry.&Introduced\\ \hline 
Kids Off Social Media Act\cite{UScongress}& Social Media providers must verify age, prohibit algorithmic recommendations,  require parental consent for users under the age of 18, and prohibit social media platform access to children under the age of 13.&Introduced\\ \hline 
Utah Minor Protection in Social Media Act (S.B. 194) \cite{utah194}& Social Media providers must verify age, enable maximum privacy settings, provide parental consent mechanisms, and protect minors' data.&Passed. It will be enacted from October 01, 2024\\ \hline 
Utah Social Media Amendments (H.B.464)\cite{utah464}& Parents can sue social media providers if social media causes mental health issues, limit social media algorithmic content to 3 hours, restrict access between 10:30 PM and 6:30 AM, and require parents' consent for minors.&Passed. It will be enacted from October 01, 2024\\ \hline 
 New York Stop Addictive Feeds Exploitation (SAFE) for Kids Act (Senate Bill S7694A) \cite{nysenate}& Prevent addictive feeds and limit the night use of children's social media accounts.&Passed on June, 2024\\ \hline
EARN IT Act\cite{earnitact}&   Prevention of online sexual exploitation of children, establishing the National Commission on Online Child Sexual Exploitation Prevention, and reporting requirements for social media providers.&Introduced\\ \hline 
Eyes on the Board Act\cite{US2023}& Schools must limit and monitor social media usage in school networks and devices.&Introduced\\\hline
    \end{tabular}

\end{table}

Government regulations are critical tools for a secure online experience and ensuring social media platforms adhere to fundamental safety rules for young and minor users. Regulations and laws can hold social media providers accountable, help increase transparency, and establish data protection measures. However, regulations are just one piece of the puzzle. Regulations must be part of a broader strategy to protect minors in the current technology age, including education, technological advancement, and industry-led and collaborative initiatives.

Certain regulations propose that age verification needs to be completed to access social media. However, this brings the challenge of personal data security. Social media providers must regulate how and where personally identifiable data is stored and used. This data should not be used for other purposes like data mining, machine learning, or monetization. The government must establish an enforcement body to closely monitor the implementation of legislation, with the power to take action against social media providers. Executives of social media providers need to be held personally accountable for decisions that have caused irreparable harm to society. Finally, if a provider is unable to fulfill their obligations, then the enforcement body should have the authority to initiate a mechanism such as a kill-switch, shutting down the platform nationwide.
 
\subsection{Industry-led Initiatives}
\subsubsection {Content Moderation}
Social media platforms must prioritize effective content moderation to ensure that online discourse remains respectful and safe for all users. With millions of individuals sharing their thoughts and opinions in real-time, social media enables diverse perspectives but also risks posting views that are considered offensive, harmful, or extreme by many users \cite{GenEco}. 

Artificial intelligence (AI) systems are not yet capable of making nuanced judgments about the context, intent, and cultural subtleties required for effective content moderation~\cite{duarte2017mixed}. AI algorithms rely on patterns learned from large datasets, which can lead to biases and inaccuracies when applied to real-world scenarios. Moreover, content moderation often requires empathy, creativity, and domain expertise, qualities that are difficult for AI systems to replicate. Human judgment is still essential in cases where the context is unclear or the situation requires understanding the subtleties, such as sarcasm, irony, or cultural references. While AI cannot replace human judgment entirely, it can be instrumental to human content moderators by triaging and prioritizing content based on language or context. For instance, AI algorithms can quickly analyze vast amounts of data to identify potentially problematic content, such as hate speech, violence, or explicit imagery, reducing the workload for human moderators. Additionally, AI-powered tools can detect language patterns, such as profanity, threats, or harassment, and flag them for further review by humans \cite{Gosztonyi2023}.

The power of human-AI collaboration lies in combining AI's triaging capabilities with human content moderators' expertise, creating a more effective and efficient content moderation process. This synergy begins when AI identifies potentially problematic content and flags it for human review, allowing trained moderators to assess flagged content, taking into account context, intent, and cultural nuances. Moderators then work with AI algorithms to make informed decisions, combining their expertise with the AI's insights~\cite{molina2022modertation}.

AI can analyze context clues, like surrounding text, hashtags, or user profiles, to better understand the intent behind a piece of content. This contextual understanding can help human moderators make more informed decisions about whether content violates community guidelines. AI algorithms can prioritize content based on factors like severity, audience reach, or potential harm caused, allowing human moderators to focus on the most critical cases. 

\subsubsection{Data Practices and Digital Responsibility}
The culprit behind this addiction-fueled chaos is the relentless pursuit of data-driven profits. Social media companies harvest vast amounts of user data, including location information, search queries, browsing history, and even biometric data (e.g., facial recognition), to create targeted advertisements that are tailored to our individual preferences. This data is then sold to third-party advertisers, who use it to craft persuasive messages that can sway our purchasing decisions.

Teenagers today are growing up in a world where social media is integral to their daily lives.  They use social media platforms to connect with friends, family, and influencers. However, data privacy and security concerns increase as they share more about themselves online, raising questions about long-term consequences of their digital footprints.

The repeated instances of data and privacy breaches among major organizations, including social media giants, have led to a decline in customer trust and increased concerns about online security and personal data protection \cite{Ayaburi2020}. The consequences of such breaches are far-reaching and devastating, with potential outcomes including identity theft~\cite{identity2007}, financial losses~\cite{poyraz2020cyber}, and even emotional distress~\cite{Labrecque2021data}. As millions of individuals entrust these platforms with their most intimate details, from passwords to credit card numbers, any lapse in security can have catastrophic repercussions. 

The digital advertising sector heavily depends on the use of behavioral tracking, which involves monitoring consumers' online activities to refine targeted advertising efforts \cite{Johnson2020}. Teenagers are particularly vulnerable to targeted advertising, as they may not fully understand how their online behavior is being tracked and used to influence their purchasing decisions~\cite{radesky2020digital}. The European Union's  Digital Services Act (DSA)~\cite{europarl} marks a positive step towards protecting consumers by prohibiting misleading practices and certain forms of targeted advertising, including those that target children or use sensitive data. Additionally, the DSA aims to prevent ``dark patterns'' and other deceptive tactics designed to manipulate users' decisions. The DSA serves as a model for other countries, emphasizing the need for similar regulations to safeguard consumer rights and promote transparency in the digital market.

\subsection{Educational Programs}
The responsibility for shaping responsible digital behavior extends beyond government and industry to include educational institutions, which have a unique opportunity to influence the next generation's online habits. Educational institutions such as schools can become the first step in helping navigate the complex issue of algorithm-driven social media's impact on teenagers, making it a valuable resource for those looking to support young people online~\cite{Dennen2020}.

Digital citizenship encompasses not just the technical skills needed to engage online but also the ethical and responsible behaviors that come with using digital technology, involving respectful and considerate interactions~\cite{Lynn2022}. Educational institutions can start by teaching students how to effectively use technology, including basic computer skills, online safety, and digital etiquette. This foundation is essential for students to navigate the digital world responsibly. Institutions can discuss the importance of online reputations and how to maintain a positive digital footprint through responsible social media use and self-reflection~\cite{Buchanan2018}. This helps students understand that their online actions have consequences and encourages them to think critically about their digital presence.

Educational institutions have the power to incorporate and integrate digital literacy into existing subjects, including health education, social studies, language arts and many more~\cite{roblyer2019integrating}. Educators can employ case studies in social studies classes to investigate historical and contemporary technological developments alongside their societal implications. Students can delve into how past and present societies have adapted to new technologies, exploring themes like the evolution of online expression, data protection, and the global economic effects of technological advancements. Schools can employ role-playing exercises that present hypothetical situations, allowing students to practice responding effectively to online conflicts or encountering inappropriate content \cite{LAURICELLA2020103989}. This hands-on approach enables students to develop essential skills in managing digital interactions and making responsible choices.

\section{A Glimmer of Hope: Mastodon's Promise}\label{Mastodon}

The rise of Mastodon~\cite{mastodon}, a decentralized social media platform, offers a glimmer of hope for those seeking a more open and inclusive online space. The decentralized nature of Mastodon allows instances to scale independently, built over ActivityPub which enables different servers to communicate with each other. \cite{MastodonGrat}.

Mastodon's promise is simple yet profound: it allows users to create their own communities (called ``instances'') where they can engage in open discussions without the constraints of corporate algorithms or interests. This decentralized approach encourages diversity, creativity, and individuality, making it an attractive option for those seeking a more authentic online experience. In traditional centralized platforms, moderation is typically managed from a single hub, where a coordinated team of moderators works in sync to oversee content across the entire platform. In contrast, decentralized systems like Mastodon fragment these moderation capabilities and efforts, introducing added complexity. Table \ref{tab:mastvscent} presents a side-by-side analysis of centralized social media platforms versus Mastodon, highlighting their similarities and differences. Each individual instance within the Mastodon network has its own distinct group of administrators and moderators responsible for reviewing and curating content as it flows through its unique community space. \cite{ACM16Mastodon}, \cite{MastodonRules}.  When an instance is not enforcing its community guidelines or is allowing harmful content to spread, defederation can be used to block communication with that instance. This means that users from other instances won't be able to interact with the problematic instance, reducing the spread of harmful information and mitigating its impact. \cite{Colglazier_TeBlunthuis_Shaw_2024}

Teenagers are especially well-positioned to benefit from Mastodon's community-based approach, which aims to offer a safe environment with responsible moderation, free from corporate algorithms, data collection and tracking. Such models offer young people a chance to develop healthy online habits and cultivate meaningful connections with others by providing a platform that values openness, inclusivity, and creativity.

\begin{table}[h!]
    \centering
\caption{Mastodon vs. Centralized Social Media: A Comparison of Key Features}
\label{tab:mastvscent}
    \begin{tabular}{|p{1.5cm}|p{3cm}|p{3cm}|}
        \hline
        \textbf{Feature} & \textbf{Mastodon} & \textbf{Centralized Social Media}\\ \hline
        Platform Ownership & Each instance is owned by its administrators & Owned by a single company or organization \cite{LACAVA2022100220}\\ \hline
        Content Moderation & Each instance has its own moderation team and rules& Single moderation team and rules apply across the platform\\ \hline
        Network Connectivity & Instances communicate with each other through open standards (e.g., ActivityPub)\cite{LACAVA2022100220, ACM16Mastodon} & Users connect to a single server or hub\\ \hline
        Scalability & Scalable through instance federation (multiple instances can communicate with each other)  & Scalable through infrastructure upgrades and load balancing\\ \hline
        Governance & Each instance has its own community-driven moderation and decision-making process & A single entity makes decisions about platform policies and moderation\\ \hline
    \end{tabular}

\end{table}
Mastodon provides many benefits, such as a decentralized platform for free speech and community building. However, it also has downsides, notably the potential to create echo chambers \cite{EchoChamber}. An echo chamber is an environment where people are only exposed to information and perspectives that reinforce their beliefs and biases without being challenged or confronted with opposing viewpoints. When users join a Mastodon instance, they are often drawn to communities that align with their interests, values, or demographics. This can lead to a self-selecting process where like-minded individuals congregate in specific instances, creating an environment where similar perspectives and ideas dominate \cite{HOBOLT_LAWALL_TILLEY_2023}. As Mastodon instances grow, users may curate content by favoriting, reblogging, or commenting on posts that align with their beliefs. This can create a feedback loop where only certain types of content are amplified, making it difficult for opposing viewpoints to gain traction. Mastodon lacks powerful automatic content moderation and is highly dependent on community efforts for self-moderation. This means that users are responsible for reporting and addressing any issues with content rather than relying on AI-powered moderators. 

Mastodon's model is a step in the right direction towards creating a healthier, more inclusive and decentralized online ecosystem. Although this approach informs potential improvements to traditional social media platforms, it is not intended as a direct replacement, but rather serves as a complementary platform. By leveraging Mastodon's strengths and understanding its limitations, we can work towards creating a comprehensive, informed technology solution that addresses the root causes of social media's problems.

\section{A Call to Action}\label{CallToAction}
Based on our findings, we propose the following multi-pronged call to action:

\begin{enumerate}

\item \textit{Strengthen Legislation and Enforcement Bodies}: Government and social media companies must provide transparent and age-appropriate content guidelines. Social media providers must regulate algorithmic transparency and make regular updates available to the public. Adopting Mastodon's decentralized approach allows each instance to enforce its own public moderation policies, ensuring users understand content guidelines and promoting open-source algorithm inspection.

\item \textit{Integrate Digital Wellness Education}: Educational institutions must incorporate digital wellness education into their curriculum, focusing on psychological impacts and enabling students to develop essential skills in managing digital interactions and making responsible choices.

\item \textit{Regularly Review and Update Policies}: Conduct regular data collection and processing audits to ensure compliance with evolving regulations (e.g., GDPR, CCPA).

\item \textit{Prioritize Teenagers' Well-being}: It is essential that social media companies implement robust measures to prevent the sharing of explicit content, especially among teenagers. This involves not only flagging inappropriate material but also educating users about online safety and responsible behavior.

\item \textit{Mitigate Algorithmic Bias through Transparent Governance}: Other social media providers can emulate Mastodon's transparent governance model to reduce algorithmic bias effectively. This involves enabling open discussions and scrutiny of the decision-making process.
\end{enumerate}

\section{Conclusion}\label{Conclusion}

This paper calls for action from the government, social media providers, and educators to work collaboratively to protect our young generation and solve social media's psychological impact. Government legislation plays a critical role in establishing the regulatory framework to establish safety rules for social media providers to protect children online. However, the enforcement of the legislation must be established. The phenomenon of doom scrolling exacerbates these mental health issues, creating a cycle of harmful content consumption that is particularly detrimental to younger, impressionable minds. The social media industry must collaborate and take initiatives such as enhanced content moderation and algorithmic adjustments. In addition, social media providers must have children-friendly safety features and bring greater accountability for their platforms.

Furthermore, exploring alternative platforms, such as Mastodon, offers a glimpse into the potential of decentralized social media models that prioritize user control and privacy. While these platforms have challenges, they represent a valuable direction for future research and development. In summary, addressing the impact of social media on children demands an integrated approach, combining legislative action, industry responsibility, educational initiatives, and continued innovation in social media platforms.

\section*{Acknowledgment}

Grammarly was utilized in the writing process to assist with editing, spell checking and grammar enhancement. All content and ideas presented in this paper are our own.

\printbibliography

\end{document}